\documentclass[runningheads,a4paper,10pt]{llncs}
\pdfoutput=1
\usepackage[utf8]{inputenc}
\usepackage{amssymb}
\usepackage{graphicx}
\usepackage{amsmath}
\usepackage{url}
\newcommand{\keywords}[1]{\par\addvspace\baselineskip
\noindent\keywordname\enspace\ignorespaces#1}

\usepackage{array}
\usepackage{multirow}
\usepackage{multicol}
\usepackage[table]{xcolor}
\usepackage{tikz,tikz-qtree,tikz-qtree-compat}
\usetikzlibrary{calc}
\newcommand{\head}[1]{\textnormal{\textbf{#1}}}

\begin{document}

\title{Visualizing BACnet Data to Facilitate Humans in Building-Security
    Decision-Making\thanks{\em The final publication is available at link.springer.com}}

\titlerunning{Visualizing BACnet Data}

\author{%
    Jernej Tonejc\inst{1}
    \and Jaspreet Kaur\inst{1,2}
    \and Adrian Karsten\inst{1,2}
    \and Steffen Wendzel\inst{1}
}

\authorrunning{Tonejc et al.}

\institute{Fraunhofer FKIE, Bonn, Germany\\
\email{\{jernej.tonejc, steffen.wendzel\}@fkie.fraunhofer.de}
\and
Rheinische Friedrich-Wilhelms-Universität Bonn, Bonn, Germany\\
\email{\{kaur, karstena\}@informatik.uni-bonn.de}
}

\maketitle
\vspace*{-5mm}

\begin{abstract}
Building automation systems (BAS) are interlinked networks
of hardware and software, which monitor and control events in the buildings.
One of the data communication protocols used in BAS is Building Automation
and Control networking protocol (BACnet) which is an internationally adopted
ISO standard for the communication between BAS devices. Although BAS focus on
providing safety for inhabitants, decreasing the energy consumption of
buildings and reducing their operational cost, their security suffers due
to the inherent complexity of the modern day systems. The issues such as
monitoring of BAS effectively present a significant challenge, i.e., BAS
operators generally possess only partial situation awareness. Especially in large and inter-connected
buildings, the operators face the challenge of spotting meaningful incidents
within large amounts of simultaneously occurring events, causing the
anomalies in the BAS network to go unobserved. In this paper, we present the
techniques to analyze and visualize the data for several events from BAS
devices in a way that determines the potential importance of such
unusual events and helps with the building-security decision making. We
implemented these techniques as a mobile (\textit{Android}) based
application for displaying application data and as tools to analyze the
communication flows using directed graphs.

\keywords{BACnet, building automation, visualization, data analysis, directed graphs, treemaps}

\end{abstract}

\section{Introduction}
\label{section:1}

Building Automation Systems (BAS) aim at controlling, monitoring and
administrating services such as heating, ventilation, air-conditioning and
lighting in the buildings. While managing various building systems,
they ensure the operational performance of the facility as well as the
comfort and safety of the building's inhabitants. They also aim to decrease
the energy consumption and reduce the operational costs of a building.

One of the data communication protocols used in BAS is Building Automation
and Control networking protocol (BACnet) \cite{ISO16484}. It is an
internationally adopted ISO standard for the communication between BAS
devices and implemented in products by more than 800 vendors worldwide. The BACnet
protocol defines a number of services that are used to communicate between
building devices and works over a number of data link/physical layers,
including ARCNET, Ethernet, BACnet/IP, Point-To-Point,
Master-Slave/Token-Passing, LonTalk etc.\ as shown in
Fig.~\ref{fig:bacnetosi}.

\begin{figure}[h]
    \centering
    \renewcommand{\arraystretch}{1.3}
    \setlength{\tabcolsep}{4pt}
    {\sffamily
    \begin{tabular}{|l|c|c|c|c|c|c|c|}
        \multicolumn{1}{l}{\bfseries OSI Layer} & \multicolumn{7}{c}{\bfseries BACnet Stack Protocol} \\\hline
        Application &  \multicolumn{7}{|c|}{BACnet Application Layer} \\\hline
        Network &  \multicolumn{7}{|c|}{BACnet Network Layer} \\\hline
        Data Link & \multicolumn{2}{|c|}{$\qquad$\parbox[c][1cm][c]{2.5cm}{BACnet/IP over\\ ISO 8802-2 LLC} } &
        MS/TP & \multirow{2}{*}{LONTalk} & PTP & BVLL & BZLL \\\cline{1-4}\cline{6-8}
    Physical & \multicolumn{1}{|c|}{Ethernet} & ARCNET & RS485 & & RS232 & UDP/IP & ZigBee \\\hline
    \end{tabular}
    }
    \renewcommand{\arraystretch}{1}
	\caption{BACnet OSI layers as defined in the ISO standard \cite{ISO16484}.}
	\label{fig:bacnetosi}
\end{figure}

BAS are responsible for taking care of many services in the buildings but
the BAS operators face a significant challenge while effectively monitoring BAS due to
modern-day building's complexity. BAS operators generally possess only
partial situation awareness, i.e., the perception of the \emph{current
situation} within the building and how it might change in the near future.
Especially in large and inter-connected buildings, the operators
face the challenge of spotting meaningful incidents within large amounts of
simultaneously occurring events.  For example, slight temperature changes
can occur throughout the day, and even hardware failures and access events
for rooms can occur hundreds of times each day. On the other hand, if a
usually closed window is opened at night, it should raise the attention of
the BAS operator: opening a window at night is probably linked to a more
important cause than slight temperature changes over the day.

In this paper, we present the techniques to analyze and visualize the data
for several events from BAS devices in a way that determines the potential
importance of such unusual events and helps with the building-security
decision making. First, we explain the concept of processing and selecting
the data from BACnet traffic to carry out the data analysis. Second, we
discuss the visualization methods that can be implemented in BAS to increase
situation awareness among the operators and users, while handling the events
effectively. These methods can improve the detection of the anomalies in
BAS. We implemented these techniques as a mobile (\textit{Android}) based
application to efficiently visualize the application data and created tools
for visualizing and analyzing message flows between different devices in
a BAS network. The tools generate directed graphs that illustrate the network
topology, allowing the operators to quickly identify unusual communication.

The rest of the paper is structured as follows. In Sect. 2, we summarize the
related work in the field of network analysis and visualization techniques for BAS. In
Sect. 3, we explain the methods used for collecting the BAS data. Section 4 discusses the visualization techniques for network data. This is 
followed by the discussion on techniques for visualizing application data along with the usability study in Sect. 5. We present the conclusions and future work in Sect. 6.

\section{Related work}
\label{section:2}

Traffic flow measurement has been known in IP networks for quite some time,
but observing the traffic flow in BACnet networks has been done only
recently \cite{celeda-traffic}. The methods were later improved in
\cite{celeda-flow} to include entropy-based analysis of the traffic flow.
However, these papers focused on the volume of the packets and packet rates
and no attempt was made to classify the traffic between individual BACnet
devices, making the methods unsuitable for detecting anomalies within the
BACnet networks.

There has been some research on visualizing events in BAS. Wendzel et
al.~proposed one such approach for visualization of simultaneously occurring
events in \cite{wendzel1} and implemented it in a tool named Chronos
\cite{chronos}. Chronos provides temporal mosaic charts for events which
provide good representation of details and make efficient use of screen
real-estate, unlike Gantt charts. The temporal mosaic charts possess the
capability of combining parallel occurring events into a single stream so as
to use the provided screen space more efficiently. In \cite{wendzel1}, the
authors make use of entropy to highlight the important events. However,
while a lot of research has been done in the context of visualizations on
limited screen space, no attempts have been made to transfer this to the
context of mobile applications in the area of BAS networking data.

Chittaro \cite{Chittaro} investigated general aspects of visualizations on mobile
devices, stressing the importance of data selection, efficiency of space
usage and interactivity.
Games et al.~\cite{games} proposed focus-plus-context approach, showing that
it outperforms bar graphs or scatter plots in certain scenarios. Our approach on
visualizing BAS data on mobile devices originates from a method called
\emph{treemaps}, first introduced by Shneiderman in 1992
\cite{Shneiderman}.  It maps a tree to rectangles for each node, filling the
available space. The rectangle sizes can be connected to some attributes of
the nodes, e.g., weights. This technique is more space efficient than
conventional graph visualizations.

\section{Collecting the BAS Data}
\label{section:3}

In this section, we describe the experimental lab setup,
our methods of obtaining the BAS data, and the sources of BAS
data that were used to obtain the results.

\subsection{Lab setup}

\begin{figure}
    \centering
    {\sffamily
        \includegraphics{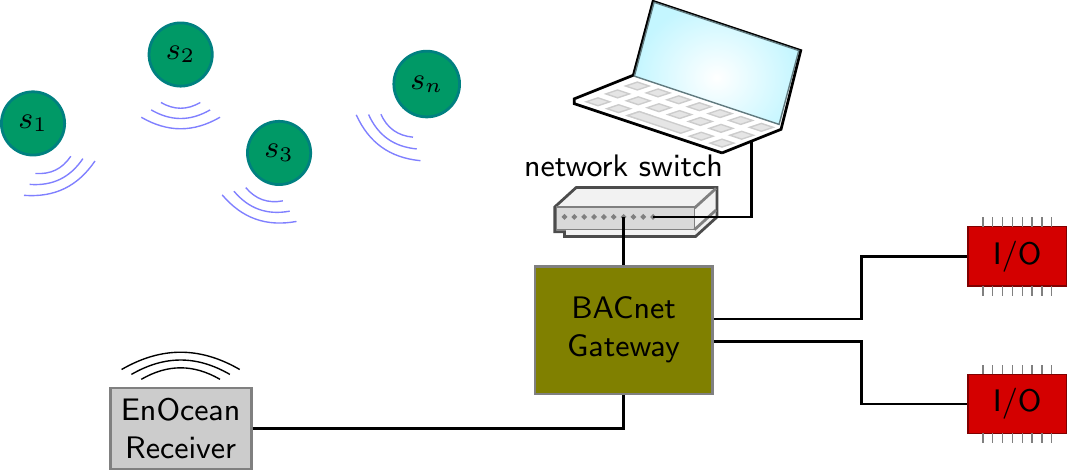}
    }

    \caption{A generalized scheme of our BACnet labs. EnOcean sensors
    $s_1,\dots,s_n$ connect wirelessly to the EnOcean receiver which is
connected to a BACnet gateway device. The binary input/output devices are
connected to the gateway via MS/TP.}
    \label{fig:lab}
\end{figure}

To evaluate our methods, we used two BACnet labs, each with several BACnet
devices. A generalized scheme of the labs is shown in Fig.~\ref{fig:lab}.

The first lab setup has 16 wireless EnOcean sensors, which includes door
and window sensors, motion sensors (wall and ceiling) and temperature
sensors. It also has 16 actuators/binary output devices. All the devices
are connected to the BACnet network via a BACnet gateway module. The second
lab setup has various BACnet devices, sensors (CO$_2$, light, motion,
presence, air pressure, moisture, temperature) on MS/TP bus, together with
32 binary outputs.

\subsection{Collection methods}

We analyzed two types of data: network data and application data. The
methods of obtaining data differ between the two types. For collecting the
application data from BAS devices we used the log export functions of the
BACnet gateway module. Module's internal software maintains the logs of data
and allows them to be exported in the form of \texttt{csv} files for each
individual sensor. In general, there are two ways of obtaining sensor data:
by periodically polling the sensors for data or by subscribing to sensor's
notifications about changes in value. We use the latter approach, i.e., the
sensors were configured to report the \textit{Change-of-Value} data.
The door, window and motion sensors record their state using binary
values, while the temperature sensor records every change in the room
temperature of more than 0.1$^o$C. The states of the actuators are also
recorded as binary values. The values from different sensors can be
combined together and correlated to analyze and visualize the events in BAS.

We used \texttt{Wireshark} \cite{wireshark} to capture the network data. Wireshark is an
open-source network protocol analyzer. It allows examining the data from a live
network or from a saved capture file. The captured data can be
interactively browsed, delving down into all the levels of packet details.
It supports hundreds of protocols and media types including BACnet.
Wireshark also has the functionality of filtering the captured packets. In
case of BACnet, packets can be filtered by using keywords
\emph{bacnet}, \emph{bvlc} etc. The captured packets are saved in .pcap
format.
In addition to our own \texttt{Wireshark} recordings from the two labs,
we used the network traffic recordings from Steve Karg's collection
\cite{kargs} to test the performance of our code.

\section{Visualizing the Network Data}
\label{section:4}

In this section, we focus on the network-related data, more specifically, on
BACnet/IP over Ethernet. In this case, the BACnet data is encapsulated in
User Datagram Protocol (UDP) layer. BACnet data is transmitted as packets
and each packet contains communication and control values. BACnet
standard requires that there exists exactly one message path between any two
nodes on an interconnected BACnet network. Therefore, a \textit{flow} between
two BACnet devices is well-defined and can be identified by the BACnet
addresses of the devices in question.

\subsection{Selection and processing}

\begin{figure}
    \centering
        \includegraphics{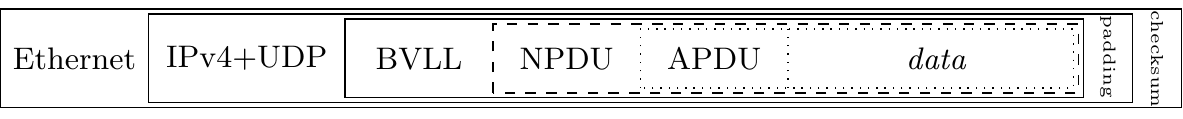}
    \caption{Structure of a typical BACnet/IP packet. The outermost layer is
        the Ethernet layer, which encapsulates the Internet Protocol (IP)
        and User Datagram Protocol (UDP) layers. BACnet/IP is encapsulated
        within the UDP datagram as a BACnet Virtual Link Layer (BVLL),
        containing the BACnet Virtual Link Control (BVLC) that indicates
        the function of the BACnet packet. The presence of the Network layer
        Protocol Data Unit (NPDU) and the Application layer Protocol Data
        Unit (APDU) within NPDU depends on the particular BVLC function and
        the values in the Network Protocol Control Information (NPCI), which
    is the second octet of NPDU.}
    \label{fig:BACnet}
\end{figure}

An important aspect of any raw data analysis is a selection of the relevant
features, together with pre- and postprocessing. Since the data are captured
using \texttt{Wireshark} and stored in a raw \texttt{pcap} format, a fair amount of
preprocessing is needed. The preprocessing was done using the Python library
Scapy \cite{scapy}, together with the open source Scapy BACnet extension
\cite{scapy-bacnet} which we substantially extended to enable full network
and application layer parsing.
Figure \ref{fig:BACnet} shows the structure of a typical BACnet/IP packet.
For a detailed explanation of the structure of NPDU and APDU, we refer the
reader to the ISO standard \cite{ISO16484}.

One piece of data that is important for analysis but is not explicitly contained
in any of the layers is the timestamp of the captured packet. However, this is
recorded by \texttt{Wireshark} and is stored for each packet along with the
packet contents inside the \texttt{pcap} capture files.
We use the timestamps to estimate the packet rates and
determine whether certain types of packets are periodic or sporadic, e.g.,
network layer message packets with a specific message type or application
layer packets with specific PDU type for a given pair of source/destination
addresses.

We next describe the features we select from each layer, starting with the outermost layer and working towards the inner layers.

\subsubsection{Ethernet layer.}
Two features are of interest here: \textit{destination MAC}
and \textit{source MAC}. These are the physical addresses of the
communicating devices, or in the case of a broadcast message,
\texttt{FF:FF:FF:FF:FF:FF} (only as destination).

\subsubsection{BVLC layer.}
Although this layer can contain broadcast distribution table data, we only
focus on the \textit{BVLC function} field and the total \textit{BVLC
length}. Both serve the purpose of characterizing the message flows.

\medskip\noindent
The last two layers, NPDU and APDU, are present and
can be analyzed only if the BVLC function is \texttt{0x04}
(\textit{forwarded NPDU}), \texttt{0x0A} (\textit{original unicast NPDU}),
or \texttt{0x0B} (\textit{original broadcast NPDU}).

\subsubsection{NPDU layer.}
The fields in NPDU layer depend on the values of the NPCI control octet,
which is the second octet within the NPDU. In particular, bit 7 of the NPCI
control octet indicates whether the NPDU contains a network layer message or
application layer data. We extract the \textit{source address},
\textit{destination address}, and \textit{message type}, when they are present.
For BACnet/IP, the source and destination addresses are always 6 octets long
and are encoded according to Annex J of the standard \cite{ISO16484} as
4-octet IP address and 2-octet port number. For MS/TP, the addresses are 1
octet long.  The addresses are needed to map the flows within the BACnet
network.

\subsubsection{APDU layer.}
If the indication bit of the NPCI control octet is 0, the data portion of
the NPDU contains application data within an APDU, from which we only
consider one parameter, the \textit{PDU type}.

The chosen fields allow us to detect and map message flows, characterize
these flows, and create an overview of the communication within a BAS
network. Characterizing individual flows also provides a means to detect
certain traffic anomalies and attacks. For example, the packet length for a
specific PDU type usually does not deviate much from the average size, so
unusually small or large packets can indicate an attack or other kinds of
anomalies.

In the postprocessing step we aggregate and export the data in a form
suitable for visualizing the flows and creating the model for anomaly detection.

\subsection{Analysis}

To analyze the flow data, we group the packets based on their source and
destination addresses and call each such group a \textit{connection}. For
each such connection, we divide the packets into two sets: packets
containing network layer messages and packets containing application layer data.
Within each set we further group the packets based on the network message
type and PDU type (for network layer messages, respectively application
data). For each set we analyze the timing data to detect whether the packets are
sent periodically or sporadically. We detect this by computing the average
inter-arrival time $\tau$ between the packets and its standard deviation. If the
standard deviation is suitably small (comparable to a fraction of the mean,
e.g., less than 20\%), we classify such traffic as \textit{periodic}. If the
standard deviation is comparable to the mean (e.g., between $0.5 \tau$ and
$2\tau$), we treat such traffic as \textit{sporadic} and we model it using a
Poisson process with parameter $\lambda = \frac{1}{\tau}$. If the standard
deviation does not fit in these two cases, we do not attempt to classify the
packets based on their arrival time.

After computing these parameters for all connections from the sample traffic
data, we create a probabilistic flow map, which we later use to analyze
each incoming packet by checking whether it fits
in the constructed flow map. We do this by computing the likelihood of
observing such a packet at the given time. If the probability of observing
such a packet is smaller than some predefined threshold (which can be
connection-dependant), we flag such packet as anomalous.

\subsection{Directed graphs for flow data}

\begin{figure}
    \centering
    \includegraphics[width=\textwidth]{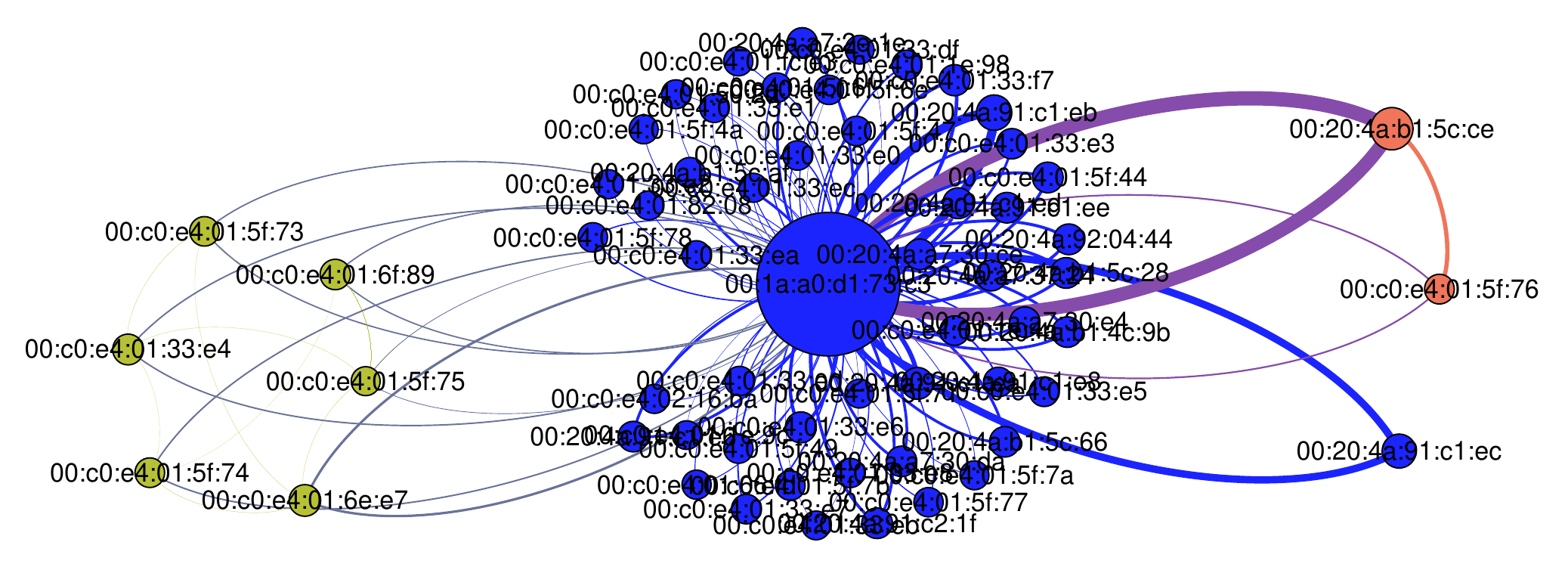}
    \caption{An example of a flow map for BACnet network using the
        application layer data.
        The thickness of the edges is determined by the weight, which
    corresponds to the fraction of the traffic belonging to that edge.}
    \label{fig:flows}
\end{figure}

In addition to analyzing the flow data, we visualize the connections between
the BACnet devices in the form of a directed graph. Each source and each
destination address represents a node, with edges being the observed
directed message flows. Both network layer messages and application layer
data can be used for this pupose and lead in general to different graphs.
The weights of the edges are computed from the number of packets on each
connection as a fraction of the total traffic. The data are processed using
a Python script and the graph information is exported in the \textit{Graph
Exchange XML Format} (GEXF).  We then visualize this flow map using the open
source program \texttt{GePhi} \cite{ICWSM09154}.  The program has several
tools to process graph data, including optimal node placement based on edge
weights and node degrees, finding clusters of more tightly connected nodes,
etc. This allows us to visualize the intrinsic network features of a BAS. An
example of a flow map is shown in Fig.~\ref{fig:flows}. The corresponding
flow classification (the top 5 flows) is shown in Table~\ref{tab:flows}.

\begin{table}
    \centering
    \caption{Flow classification of the 5 most frequent flows for the
        network in Fig.~\ref{fig:flows}. The PDU types are
        \texttt{0x00}-BACnet-Confirmed-Request and
        \texttt{0x03}-BACnet-ComplexACK.}
    \label{tab:flows}
    \begin{tabular}{>{\centering\arraybackslash}p{1.9cm} >{\centering\arraybackslash}p{1.9cm} >{\centering\arraybackslash}p{1.9cm} >{\centering\arraybackslash}p{1.9cm} >{\centering\arraybackslash}p{1.9cm} >{\centering\arraybackslash}p{1.9cm}}\hline

        \head{Source} & \head{Destination} & \head{PDU type} & \head{$\tau$} & \head{$\sigma$} & \head{Flow type} \\\hline

        \texttt{73:c3} & \texttt{5c:ce} & \texttt{0x00} & 0.96743 & 1.75864 & sporadic\\
        \texttt{5c:ce} & \texttt{73:c3} & \texttt{0x03} & 1.02827 & 1.88999 & sporadic\\
        \texttt{73:c3} & \texttt{c1:eb} & \texttt{0x00} & 1.48328 & 2.93323 & sporadic\\
        \texttt{c1:eb} & \texttt{73:c3} & \texttt{0x03} &  1.48876 & 2.97395 & sporadic\\
        \texttt{73:c3} & \texttt{5f:44} & \texttt{0x03} & 60.9053 & 0.07921 & periodic\\\hline
    \end{tabular}
\end{table}

We construct the directed graph representation from traffic recordings for a
sample period, e.g., one week. We then regenerate the representation every
day by adding the newly recorded data from that day and observing the
changes in the graph. If new nodes or edges appear, the operator is
notified. The new nodes and edges can be confirmed and become part of the
reference representation from that point on. Additionally, packets can also
be analyzed in real time by checking whether they fit the individual
connection distribution and timing parameters.

\section{Visualizing the Application Data}
\label{section:5}

The application-related data are the actual sensor values and actuator
states. No special selection is performed on the application data, we simply
collect all the values that are available. Since the application data come
from internal logs, no preprocessing is needed on our part. For the
postprocessing, we convert the extracted \texttt{csv} data from internal logs in a format that is more
appropriate for our visualization application. In particular, since most
sensors report Change-of-Value data and our visualization methods are based
on fixed time intervals, we extrapolate the data so that there is at least
one data point within each 15 minute interval, by repeating the previous known
sensor value.

\subsection{Weighted tree for application data}
\label{subsection:5.1}
\begin{figure}
    \centering
    {\sffamily
        \includegraphics{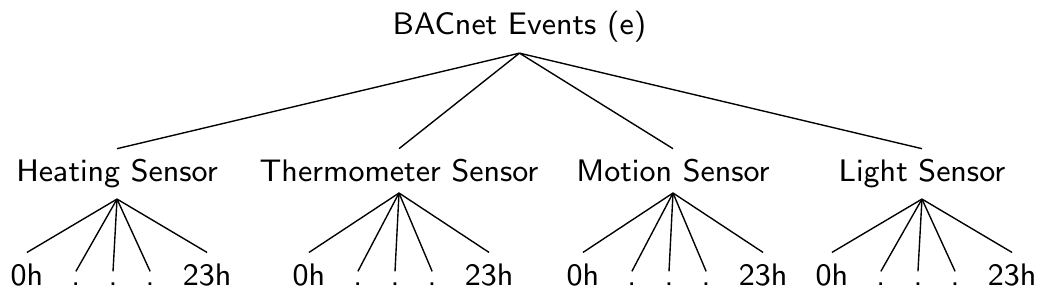}
    }
    \caption{Weighted tree for BACnet application data. Events for each day are
    clustered by sensor type and subdivided in individual hours within each cluster.}
	\label{fig:weightedtree}
\end{figure}

The BACnet application data is transformed into a weighted tree for each day as shown in
Fig.~\ref{fig:weightedtree}. The root represents the total set of BACnet
events on that day, while its children separate the data into a specific
cluster for each sensor type. Each cluster vertex has 24 children,
representing the 24 hours of a day. In \cite{wendzel1}, the authors already
considered entropy for highlighting events in BAS. We expand on their
work as follows. The weight $W(h)$ for an hour $h$ is a
function of the information content $I(h)$ and the number of changes in
value $N(h)$ in that hour. The events which deviate in value from the events
in the past are associated with higher weights. The calculation differentiates
between floating point and boolean values. For a boolean event $e$ with
value $v$, the probability $p(v)$ is calculated by comparing it to
the values of events in the past at around the same time of the day.

The information content $I(e)$ is computed as
$$
    I(e) = -\log_2(p(v)).
$$
For a floating point event $e$, the deviation of its value $v$ from the mean of the
values in the past at around the same time is considered. $I(h)$ results from
the maximum of $I(e)$ in hour $h$. The number of changes in value in hour $h$ is also
compared to the number of changes in the past and $N(h)$ is the deviation from
the resulting mean. Finally, $W(h)$ is the maximum of $I(h)$ and $N(h)$. These
calculations result in weights that are higher for data whose fluctuation
deviates significantly from the average or whose values are unusual for the
specific time. However, since real BACnet traffic will always show some form
of irregularity or inconsistency which is not caused by malfunctions, the
weight calculation will generally produce noise which might lead to false
conclusions. The weight for a cluster vertex $W(c)$ is the mean of the weights
$W(h)$ of its 24 children.

\begin{figure}
    \centering
    \includegraphics[scale=0.2]{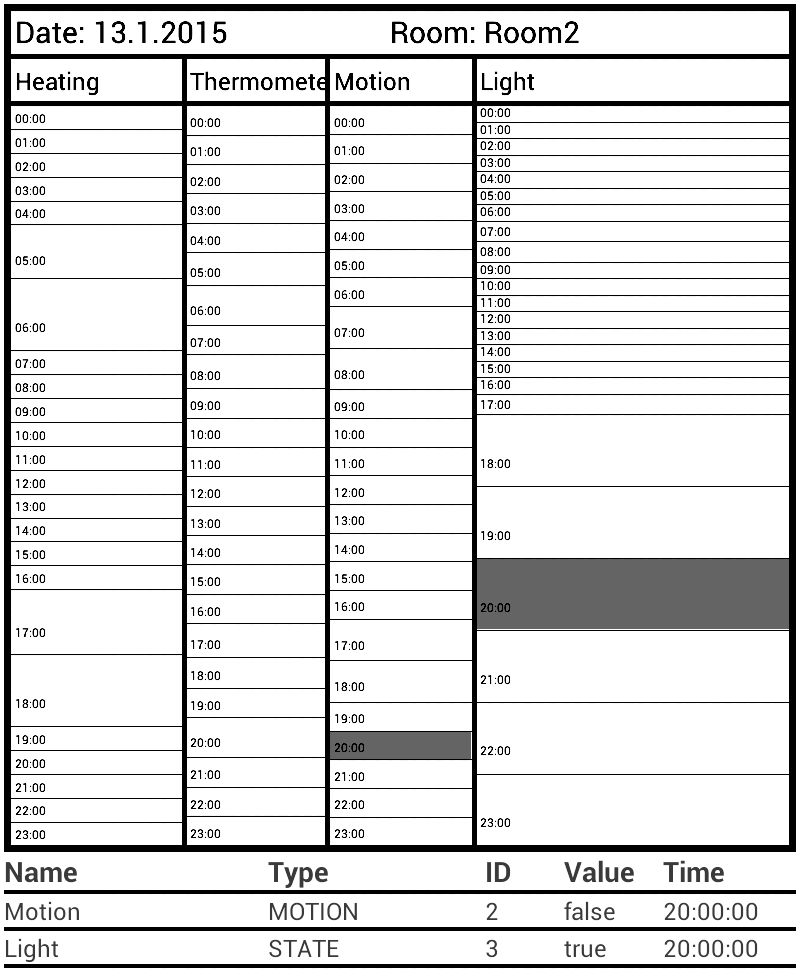} $\qquad$
	\includegraphics[scale=0.2]{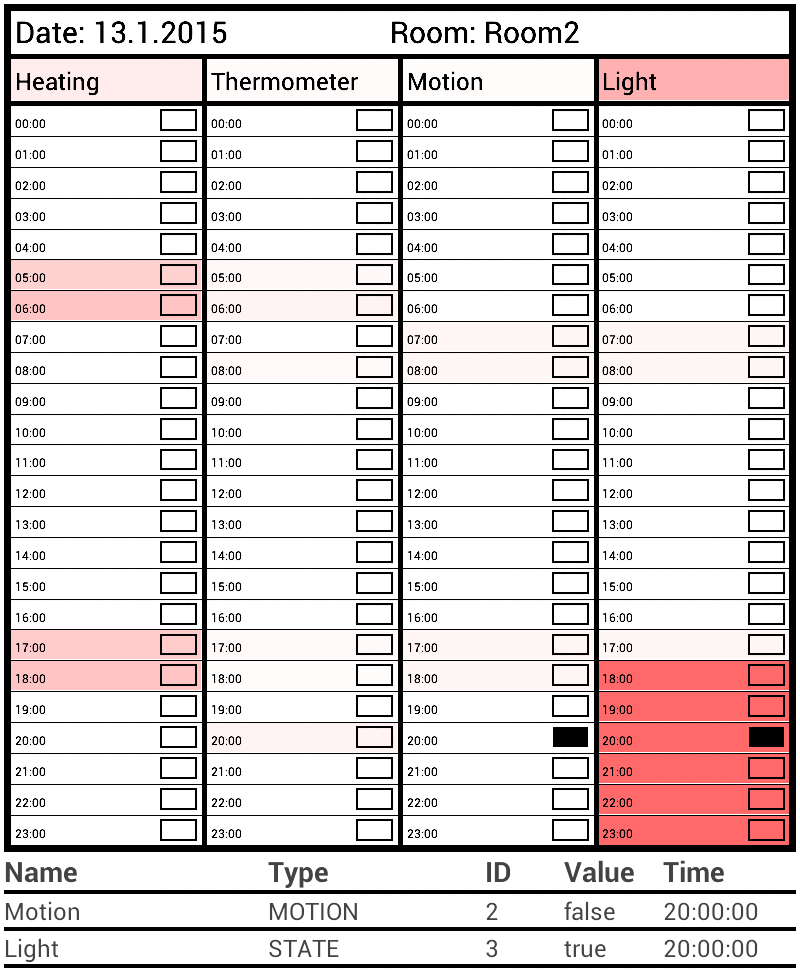}
    \caption{The size coding method (\textit{left}) and the color coding method
    (\textit{right}).}

	\label{fig:screenshots}
\end{figure}

The visualization of the weighted tree, representing one day of data, is
implemented using a 2D-grid. Figure~\ref{fig:screenshots} shows two screenshots of our
implementation. Every column is associated with a cluster of sensors and
there is a row for each of the 24 children. The details of data events associated
with a row can be expanded separately for every cluster and viewed in a list
below the grid. We implemented two methods for the visualization of the
weights, namely, the \textit{size coding} and the \textit{color coding}.
Both methods are illustrated in Fig.~\ref{fig:screenshots}. In the
size coding method, we link the weights to the heights of the individual cells
and the widths of the whole columns. This results in bigger rectangles for interesting
data. In the color coding method, we use uniform grid size while
visualizing the weights using colors. The background of a row is
linked to the weight of that row, giving it darker shades of color for
higher weights. The weights for the clusters are visualized by colored
backgrounds of the column headings.
We implemented both visualization methods of the BACnet application data for
Android phones. Due to the limited display space available on these devices,
we focused on a space efficient arrangement of the data, providing a good
display of values in the grid along with the feature to go into details for
each cell value. The potentially interesting data are highlighted (either by
using the size coding method or the color coding method, as shown in
Fig.~\ref{fig:screenshots}) in the grid to make the detection of anomalies
easier.

\subsection{Usability study}
We conducted a user study with ten participants to test our Android
implementations. The participants had background in the field of Computer Science
but had no deeper knowledge or practical experience in operating building
management systems. The goal was to compare the two highlighting methods (as
presented in Sect.~\ref{subsection:5.1}) for interesting data and to prove that the
visualizations enable fast and reliable error detection while being
restricted to a small screen space.

We simulated data for two typical scenarios of malfunctions in a BAS. The
first scenario represented a \emph{light} that usually gets switched on or
off in connection with the measurements of a motion sensor. From a certain
point onward the light stayed switched on all the time, even if the motion
sensor did not detect any motion because the automatic regulation was
broken. In the second scenario, a \emph{thermometer} got stuck at a certain
value and did not measure the real temperature any more. This caused the
heating to detect that the room was warm enough and so it turned itself off
automatically.

The participants were split into two groups of five each. The first group
was asked to analyze the first scenario using the color coding method and the
second scenario using the size coding method as highlights for unusual
information. The second group was asked to do the same but the order of
scenarios was reversed. All the participants were also asked to fill out a
questionnaire which contained the questions regarding their confidence in
the correctness of their answers, their opinion about the usability in
general and their preference for the highlighting method.

The total amount of time needed to handle the two scenarios was 12\% less
for color coding method in comparison to the size coding method. This result
is consistent with the
feedback on confidence, intuitiveness and efficiency of screen space usage,
which are all slightly better for the color coding method. When being asked about their
preference, eight out of ten preferred the color coding method over the size
coding method. However, with a
correctness of 72.5\%, the size coding method seems to be superior in this regard compared
to only 52.5\% correctness for the color coding method. This discrepancy can be explained
by the users relying too much on the colored highlights and neglecting to
verify the underlying data. Since there will always be some highlights which
do not correspond to actual errors in the system, neglecting the real data
can lead to false conclusions. In addition to being less obtrusive, size
coding method has an advantage that interesting rows are bigger and can be selected
easier for looking at the data list. This supports a more thorough analysis
and requires longer time but leads to better conclusions. Due to the variable row
sizes, a disadvantage of size coding method is that the rows in different columns are not
aligned, which makes the comparison across different sensors harder.

\section{Conclusions and Future Work}
\label{section:6}
Visualizing network message flows and computing probabilistic flow maps to
study the traffic patterns allows us to detect various kinds of anomalies
and attacks that could be present within a BACnet network. By focusing
solely on BACnet traffic, the amount of data is reduced, which makes it more
manageable for analysis. However, our current models support only two
time-based types of traffic, periodic and sporadic, without the capacity to
characterize diurnal, weekly or seasonal cycles which are common in BAS. We
therefore intend to incorporate these more advanced classifications of flows
in our future work to improve anomaly detection.

The usability study has shown that the visualization techniques for
application data work. 
Considering that the participants had no practical experience in the BAS
field, 72.5\% and 52.5\% correctness is a satisfying result.
The feedback in general supports this conclusion. However, there is
still some room for improvement. A combination of both highlighting methods
is possible, combining the advantages of both and reducing the disadvantages
to a minimum. In addition, further testing has to be done to improve the
highlight calculations, reduce their noise and increase their significance.

%\bibliography{references}

\end{document}